\documentclass[preprintnumbers,amsmath,amssymb]{revtex4}
\usepackage{epsfig}
\usepackage{color}
\usepackage{dsfont}
\usepackage[colorlinks,urlcolor=blue,linkcolor=blue,citecolor=red]{hyperref}
\begin{document}
\setlength{\voffset}{1.0cm}
\title{First-order phase boundaries of the massive 1+1 dimensional \\ Nambu--Jona-Lasinio model with isospin}
\author{Michael Thies\footnote{michael.thies@gravity.fau.de}}
\affiliation{Institut f\"ur  Theoretische Physik, Universit\"at Erlangen-N\"urnberg, D-91058, Erlangen, Germany}
\date{\today}

\begin{abstract}
The massive two-dimensional Nambu--Jona-Lasinio model with isospin (isoNJL) is reconsidered in the large $N_c$ limit. We continue
the exploration of its phase diagram by constructing missing first-order phase boundaries. At zero temperature, a phase boundary
in the plane of baryon and isospin chemical potentials separates the vacuum from a crystal phase. We derive it from the 
baryon spectrum of the isoNJL model which, in turn, is obtained via a numerical Hartree-Fock (HF) calculation. At finite 
temperature, a first-order phase boundary sheet is found using a thermal HF calculation. It interpolates smoothly between
the zero temperature phase boundary and the perturbative sheet. The calculations remain tractable owing to the assumption 
that the charged pion condensate vanishes. In that case, most of the calculations can be done with methods developed in the past 
for solving the massive one-flavor NJL model. 
 
\end{abstract}

%\pacs{}
\maketitle

%<<<<<<<<<<<<<<<<<<<<<<<<<<<<<<<<<<<<<<<<<<<<<<<<<<<<<<<<<<<<<<<<<<<<<<<<<<<<<<<<<<<<<<<<<<<< <<<<<<<<<<<<<<<<<<<<<<<<<<<<<
%<<<<<<<<<<<<<<<<<<<<<<<<<<<<<<<<<<<<<<<<<<<<<<<<<<<<<<<<<<<<<<<<<<<<<<<<<<<<<<<<<<<<<<<<<<<<<<<<<<<<<<<<<<<<<<<<<<<<<<<<<<
\section{Introduction}
\label{sect1}
%<<<<<<<<<<<<<<<<<<<<<<<<<<<<<<<<<<<<<<<<<<<<<<<<<<<<<<<<<<<<<<<<<<<<<<<<<<<<<<<<<<<<<<<<<<<<<<<<<<<<<<<<<<<<<<<<<<<<<<<<<<
%<<<<<<<<<<<<<<<<<<<<<<<<<<<<<<<<<<<<<<<<<<<<<<<<<<<<<<<<<<<<<<<<<<<<<<<<<<<<<<<<<<<<<<<<<<<<<<<<<<<<<<<<<<<<<<<<<<<<<<<<<<
Gross-Neveu models in 1+1 dimensions are fascinating as they show explicitly how a rich variety of phenomena unfolds from an extremely simple
Lagrangian, a trademark of quantum field theories. Apart from their pedagogical value as exactly solvable field theoretic models, they have found
useful applications in condensed matter, nuclear and particle physics. Originally, two variants with scalar-scalar four-fermion interactions
were considered differing in their respective chiral symmetry. We denote by Gross-Neveu (GN) model \cite{L1} the version with Z$_2\times$Z$_2$ chiral symmetry
($\psi \to \gamma_5 \psi$),
\begin{equation}
{\cal L}_{\rm GN} = \bar{\psi} i \partial \!\!\!/ \psi + \frac{g^2}{2} \left(\bar{\psi}\psi \right)^2,
\label{1}
\end{equation}
and by Nambu--Jona-Lasinio (NJL) model \cite{L2} the version with U(1)$\times$U(1) chiral symmetry ($\psi \to \exp\{i (\alpha + \beta \gamma_5)\}\psi$),
\begin{equation}
{\cal L}_{\rm NJL} = \bar{\psi} i \partial \!\!\!/ \psi + \frac{g^2}{2}\left[ (\bar{\psi}\psi)^2 + ( \bar{\psi} i \gamma_5 \psi )^2 \right].
\label{2}
\end{equation}
Both models are endowed with U($N_c$) flavor symmetry (often called ``color" in this context), thus generating a useful expansion parameter, $1/N_c$.
We suppress flavor indices as usual ($\bar{\psi}\psi = \sum_{i=1}^{N_c} \bar{\psi}_i \psi_i$ etc.) to ease the notation.
Models (\ref{1}) and (\ref{2}) are very well explored, notably in the large $N_c$ limit. They display dramatic differences related in some way or other
to the appearance of a massless pseudoscalar boson in the NJL model. More recently, SU(2) isospin symmetry has been included as suggested by the 
phenomenologically relevant 3+1 dimensional NJL variant, widely used for strong interaction studies \cite{L3}. We refer to the corresponding model
as NJL model with isospin (isoNJL),
\begin{equation}
{\cal L}_{\rm isoNJL} = \bar{\psi} i \partial \!\!\!/ \psi + \frac{G^2}{2}\left[ (\bar{\psi}\psi)^2 + ( \bar{\psi} i \gamma_5 \vec{\tau} \psi )^2 \right].
\label{3}
\end{equation}
Here, the pseudoscalar interaction is in the isovector channel, hence the would-be Goldstone boson of the NJL model is turned into a triplet of pseudoscalar, isovector ``pion" fields.
Chiral symmetry gets promoted to the non-Abelian group SU(2)$\times$SU(2). 

Each one of the models (\ref{1},\ref{2},\ref{3}) can be extended by adding a bare mass term $\delta {\cal L} = - m_0 \bar{\psi}\psi$
to the Lagrangian, breaking chiral symmetry explicitly. The corresponding variants of the GN model are then called ``massive" models, although it should be kept in mind
that the fermions acquire a dynamical mass already in the chiral limit. Like the bare coupling constant, the bare mass $m_0$ is not a physical parameter. When renormalizing the theory,
the coupling constant is traded for the dynamical fermion mass $m$ setting the scale (we use units such that $m=1$ throughout this paper), whereas the bare mass gets replaced by the 
renormalization group invariant parameter $\gamma$ (called ``confinement parameter" in the condensed matter literature)
\begin{equation}
\gamma = \frac{\pi m_0}{N_c g^2} \quad ({\rm NJL,GN}), \quad \gamma = \frac{\pi m_0}{2N_c G^2} \quad ({\rm isoNJL}).
\label{4}
\end{equation}

\begin{figure}
\begin{center}
\epsfig{file=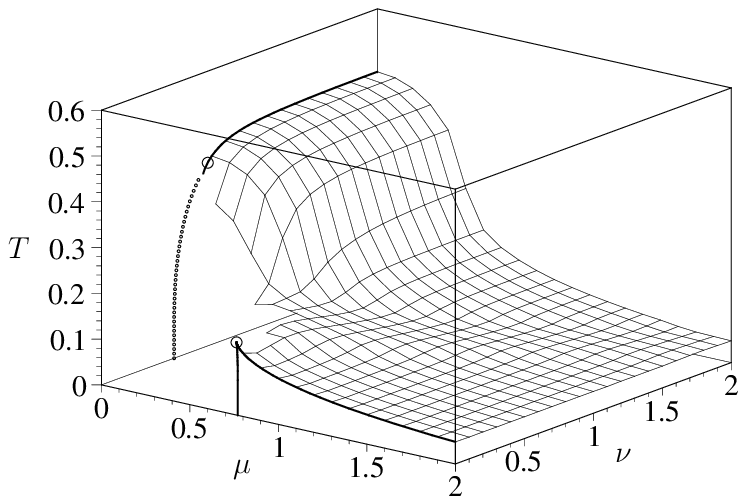,height=5.5cm,width=6.5cm,angle=0}\hskip 1.5cm\epsfig{file=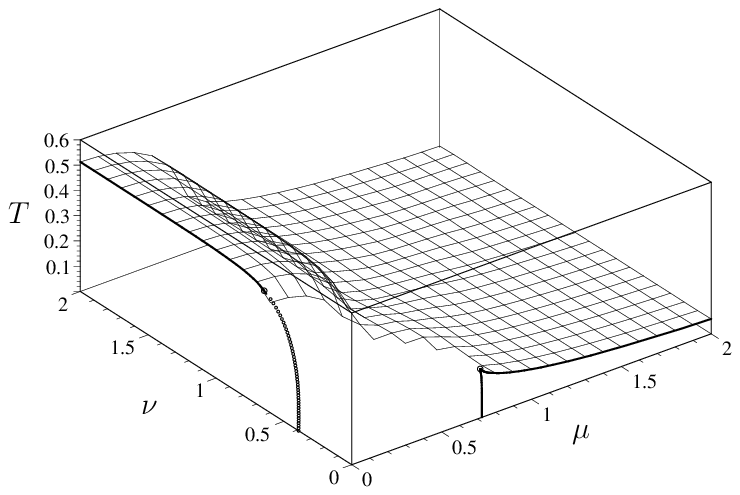,height=6.0cm,width=6.0cm,angle=0}
\caption{Present status of the phase diagram of the massive isoNJL model for $\gamma=0.1$, adapted from \cite{L15}.  A 2nd order 
sheet separates a low temperature inhomogeneous from a high temperature homogeneous phase. Two viewing angles are shown to highlight the construction site
around the origin where a first order sheet is still missing.} 
\label{fig1}
\end{center}
\end{figure}

Of particular interest for us are thermodynamic equilibrium properties and the phase diagrams of these various models. As far as the massless and massive GN and NJL models are concerned,
we have rather comprehensive information about the phase diagrams in the large $N_c$ limit, either analytical or numerical \cite{L4,L5,L6,L7,L8}. 
A prominent theme of these investigations was the identification of inhomogeneous phases, responsible for the richness of the results.
Not surprisingly, it is taking somewhat longer to map out the phase structure of the isoNJL model to the same extent. In the last few years, many partial results have been 
accumulated, typically by using variational methods or restricting the number of chemical potentials \cite{L9,L10,L11,L12,L13,L14}. Very recently, a first candidate for the full phase diagram
of the massless isoNJL model has been proposed, including all three chemical potentials $\mu,\nu,\nu_5$ (conjugate to baryon number, isospin and axial isospin) and temperature $T$ \cite{L15}.
In that work the massive model has also been addressed, although with perturbative methods only. This limits the information one can get to the high temperature phase boundary 
separating the inhomogeneous phase from the homogeneous one. To illustrate the state of the art, we reproduce in Fig.~\ref{fig1} the resulting phase diagram 
in ($\mu,\nu,T$)-space for one particular bare mass. The surface drawn is the result of a  
stability analysis and requires only 2nd order almost degenerate perturbation theory, significantly less costly than a full Hartree-Fock (HF) calculation.
As explained in \cite{L15}, the massive isoNJL phase diagram in the ($\mu,T$)-plane coincides with that of the massive GN model, whereas it matches the phase diagram of the massive NJL 
model in the ($\nu,T$)-plane (provided one identifies $\nu$ in the isoNJL model with $\mu$ in the NJL model). Fig.~\ref{fig1} strongly suggests the existence of 
a first order sheet connecting the vacuum phase in a region around the origin of the chemical potential plane with the 
perturbative sheet. Since the technique required to determine this first order sheet is nonperturbative, it had to be postponed in Ref.~\cite{L15}.
It is the purpose of the present paper to continue the construction of the isoNJL phase diagram by determining the first order phase transition sheet explicitly.
Coming back to Fig.~\ref{fig1}, what is shown there is the phase boundary sheet connecting the perturbative phase boundariy of the GN model
(in the ($\mu,T$) coordinate plane) with that of the NJL model (in the ($\nu,T)$ coordinate plane). Now we will be looking for the corresponding sheet 
connecting the nonperturbative, first order phase boundaries in the two planes, also shown in Fig.~\ref{fig1}. This is obviously needed to fully delimit the region in ($\mu,\nu,T$) space 
where the order parameter is inhomogeneous. 
 
As we shall see, it is likely that there are further first order phase boundaries inside the crystal region. These are not subject of the present work, so 
that we are not yet able to present the complete phase diagram of the isoNJL model.

%###########################################################################################################################
%####################################################################################################\#######################

This paper is organized as follows. In Sec.~\ref{sect2}, we determine the baryon spectrum of the massive isoNJL model. We find self-consistent
HF solutions closely related to baryons of the one-flavor NJL model. In Sec.~\ref{sect3}, we use the baryon spectrum to construct the phase boundary at 
$T=0$ and compare it to a homogeneous calculation. In Sec.~\ref{sect4} we complete the calculation of the phase boundary sheet enclosing the inhomogeneous phase 
by a thermal HF calculation of the 1st order sheet. Sec.~\ref{sect5} contains a brief summary of this work and an outlook. 

%<<<<<<<<<<<<<<<<<<<<<<<<<<<<<<<<<<<<<<<<<<<<<<<<<<<<<<<<<<<<<<<<<<<<<<<<<<<<<<<<<<<<<<<<<<<<<<<<<<<<<<<<<<<<<<<<<<<<<<<<<<
%<<<<<<<<<<<<<<<<<<<<<<<<<<<<<<<<<<<<<<<<<<<<<<<<<<<<<<<<<<<<<<<<<<<<<<<<<<<<<<<<<<<<<<<<<<<<<<<<<<<<<<<<<<<<<<<<<<<<<<<<<<
\section{Baryons in the massive isoNJL model}
\label{sect2}
%<<<<<<<<<<<<<<<<<<<<<<<<<<<<<<<<<<<<<<<<<<<<<<<<<<<<<<<<<<<<<<<<<<<<<<<<<<<<<<<<<<<<<<<<<<<<<<<<<<<<<<<<<<<<<<<<<<<<<<<<<<
%<<<<<<<<<<<<<<<<<<<<<<<<<<<<<<<<<<<<<<<<<<<<<<<<<<<<<<<<<<<<<<<<<<<<<<<<<<<<<<<<<<<<<<<<<<<<<<<<<<<<<<<<<<<<<<<<<<<<<<<<<<

To motivate why we are interested in baryons in the context of equilibrium thermodynamics, let us recall a simple fact known from the one-flavor GN and
NJL models. At $T=0$, there is a critical chemical potential above which translational invariance is broken and a crystal phase is favored.
This critical chemical
potential is simply given by the mass of the most stable baryon. In the present case of the massive isoNJL model, we are looking for a phase boundary 
in the plane of two chemical potentials. How is this curve related to the masses of baryons with different fermion number and isospin?
This is the subject of the present section and of Sec.~\ref{sect3}.

In the massive GN and NJL models, baryon spectra and wave functions have been computed analytically (GN) or numerically (NJL) some time ago
\cite{L16,L17,L18}. Corresponding results for the massive isoNJL model are not yet available, to the best of our knowledge. Hence our first task is to determine
the baryon spectrum. To this end, we set up the HF problem much like in Ref.~\cite{L15}, but omitting the chemical potentials. We start from the Dirac-HF
equation 
\begin{equation}
\left(-i \gamma_5 \partial_x + \gamma^0 S + i \gamma^1 \tau_a P_a   \right)\psi = \omega \psi
\label{5}
\end{equation}
supplemented by the self-consistency conditions 
\begin{eqnarray}
S-m_0 & = & -  G^2 \langle \bar{\psi} \psi \rangle ,
\nonumber  \\
P_a & = & - G^2 \langle \bar{\psi} i \gamma_5 \tau_a \psi \rangle .
\label{6}
\end{eqnarray}
Using the Dirac matrices
\begin{equation}
\gamma^0 = \sigma_1, \quad \gamma^1 = i \sigma_2, \quad \gamma_5 = \gamma^0 \gamma^1 = - \sigma_3 
\label{7}
\end{equation}
and ordering the four (iso-)spinor components as follows,
\begin{equation}
\left( \begin{array}{c} \Psi_{1,1} \\ \Psi_{1,2} \\ \Psi_{2,1} \\ \Psi_{2,2} \end{array} \right) = 
\left( \begin{array}{c} \Psi_{L,\uparrow} \\ \Psi_{L,\downarrow} \\ \Psi_{R,\uparrow} \\ \Psi_{R,\downarrow} \end{array} \right) ,
\label{8}
\end{equation}
the Hamiltonian becomes
\begin{equation}
H=
\left( \begin{array}{cccc} i \partial_x   & 0 & {\cal D}^* & {\cal C}^* \\ 0 & i \partial_x   & - {\cal C} & {\cal D} \\
{\cal D} & - {\cal C}^* & - i \partial_x     & 0 \\ {\cal C} & {\cal D}^* & 0 & - i \partial_x   \end{array} \right) .
\label{9}
\end{equation}
We have introduced two complex mean fields 
\begin{equation}
{\cal D}  = S-i P_3, \quad {\cal C} = P_2-iP_1 ,
\label{10}
\end{equation}
subject to the self-consistency conditions 
\begin{eqnarray}
{\cal D}- m_0 & = & - 2 N_c G^2 \sum \left(\Psi_{1,1}^* \Psi_{2,1} + \Psi_{2,2}^* \Psi_{1,2} \right) n_{\rm occ} ,
\nonumber \\
{\cal C} & = &  - 2 N_c G^2  \sum \left(\Psi_{1,1}^* \Psi_{2,2} - \Psi_{2,1}^* \Psi_{1,2} \right) n_{\rm occ} .
\label{11}
\end{eqnarray}
Following Ref.~\cite{L15}, we shall assume from now on that the ``charged pion condensate" ${\cal C}$  vanishes. 
The Hamiltonian (\ref{9}) then becomes block-diagonal and 
the HF equations for isospin up and down states decouple,
\begin{eqnarray}
\left( \begin{array}{cc} i \partial_x  & {\cal D}^* \\ {\cal D}  & - i \partial_x   \end{array} \right) 
\left( \begin{array}{c} \Psi_{1,1} \\ \Psi_{2,1} \end{array} \right) & = &  \omega \left( \begin{array}{c} \Psi_{1,1} \\ \Psi_{2,1} \end{array} \right) ,
\nonumber \\
\left( \begin{array}{cc} i \partial_x   & {\cal D} \\ {\cal D}^*  & - i \partial_x   \end{array} \right) 
\left( \begin{array}{c} \Psi_{1,2} \\ \Psi_{2,2} \end{array} \right) & = &  \omega \left( \begin{array}{c} \Psi_{1,2} \\ \Psi_{2,2} \end{array} \right) .
\label{12}
\end{eqnarray}
Nevertheless, the two isospin states still communicate through the self-consistency condition and the fact that their mean fields are complex conjugated.
This raises the question to what extent the solution of the one-flavor GN and NJL models can serve to find baryons of the isoNJL model.
A similar issue came up in Ref.~\cite{L15}, but now the arguments are somewhat different since there are no chemical potentials. 

We recall that a baryon in the massive NJL model has two bound states with energies $\epsilon_{1,2}$ (labeled such that $\epsilon_1<\epsilon_2$) \cite{L18}. Each level can be filled
with up to $N_c$ fermions as encoded in filling fractions $\nu_{1,2}$ (fermion number divided by $N_c$). Suppose first that we fill the bound states for up and down
``quarks" with the same filling fractions. Then ${\cal D}$ and ${\cal D}^*$ must be identical and the baryon problem reduces to that of the massive GN model (real ${\cal D}$).
On the other hand, if we fill isospin-up levels with filling fractions $\nu_1,\nu_2$ and isospin-down levels with filling fractions $(1-\nu_2),(1-\nu_1)$, respectively, we have the
correct assignment for baryon and antibaryon of the one-flavor NJL model. In that case the baryon problem reduces to that of the massive NJL model.
In these two special cases, we are through since the one-flavor baryon problems have already been solved. 
For general filling fractions, these arguments cannot be used. Surprisingly, it turns out that it is always possible to reduce the isoNJL baryon problem to the NJL baryon problem,
provided one considers not only ground states but also excited states of the one-flavor model. The way in which this works is more subtle than in
the special cases discussed so far, therefore we shall present the arguments in some detail. 

We start by spelling out the contributions to the baryon mass (divided by $N_c$) from up and down quarks for arbitrary filling fractions,
\begin{eqnarray}
\frac{M_B}{N_c} & = & E_{\rm up} + E_{\rm down} + E_{\rm d.c.},
\nonumber \\
E_{\rm up} & = & \nu_{1,{\rm up}} \epsilon_{1,{\rm up}} + \nu_{2,{\rm up}} \epsilon_{2,{\rm up}} +  \sum_{- \Lambda}^{-1} \epsilon_{\rm up} , 
\nonumber \\
E_{\rm down} & = &  \nu_{1,{\rm down}} \epsilon_{1,{\rm down}} + \nu_{2,{\rm down}} \epsilon_{2,{\rm down}} +  \sum_{- \Lambda}^{-1} \epsilon_{\rm down} .
\label{13}
\end{eqnarray}
In line with the HF approach they comprise the sum over single particle energies for all filled states (valence states and Dirac sea, regularized with cutoff $\Lambda$) and a double counting correction 
\begin{equation}
E_{\rm d.c.} = \frac{\int dx |{\cal D}-m_0|^2}{2 N_c G^2}.
\label{14}
\end{equation}
Two facts about the one-flavor NJL spectrum are important in the following. First, 
the spectrum of ${\cal D}^*$ is inverted as compared to the spectrum of ${\cal D}$. This can easily be understood by using charge conjugation. Perform a 
transformation with the hermitean, unitary matrix $i \gamma^1$. This changes the sign of $\gamma_5$ and $\gamma^0$, but not of 
$\gamma^1$. Therefore it is equivalent to changing the sign of $\gamma^1$ and of the energy eigenvalue $\epsilon$ in the HF equation (${\cal D}=S-iP$) of the one-flavor NJL model
\begin{equation}
\left( - i \gamma_5  \partial_x + \gamma^0 S + i \gamma^1 P \right) \psi = \epsilon \psi.
\label{15}
\end{equation}
Secondly, since ${\rm Tr\,} H=0$, the energy eigenvalues must satisfy 
\begin{equation}
0 = \epsilon_{1,{\rm up}} + \epsilon_{2,{\rm up}}  +\sum_{- \Lambda}^{-1} \epsilon_{\rm up}  + \sum_1^{\Lambda} \epsilon_{\rm up},
\label{16}
\end{equation}
and similarly for isospin down.
These two facts enable us to express the single particle energies for isospin down in Eq.~(\ref{13}) by those for isospin up.
Assuming $\epsilon_{1,{\rm up}}<\epsilon_{2,{\rm up}}$ we have
\begin{equation}
\epsilon_{1,{\rm down}} = - \epsilon_{2,{\rm up}}, \quad \epsilon_{2,{\rm down}} = - \epsilon_{1,{\rm up}}
\label{17}
\end{equation} 
due to the inversion of the spectrum. The continuum states yield
\begin{equation}
\sum_{- \Lambda}^{-1} \epsilon_{\rm down} =  - \sum_{1}^{\Lambda} \epsilon_{\rm up} = \epsilon_{1,{\rm up}} + \epsilon_{2,{\rm up}} + \sum_{-\Lambda}^{-1} \epsilon_{\rm up}
\label{18}
\end{equation}
where we have used (\ref{16}) in the last step. Putting everything together, we find
\begin{equation}
E_{\rm down} = \left(1- \nu_{2,{\rm down}}\right) \epsilon_{1,{\rm up}} + \left(1 - \nu_{1,{\rm down}}\right) \epsilon_{2,{\rm up}} +  \sum_{- \Lambda}^{-1} \epsilon_{\rm up}  .
\label{19}
\end{equation}
In order to determine the baryon mass in the isoNJL model, we have to minimize 
\begin{equation}
\frac{M_B}{N_c}  = \left(1 + \nu_{1,{\rm up}}-  \nu_{2,{\rm down}}\right) \epsilon_{1,{\rm up}}+ \left(1 + \nu_{2,{\rm up}}  - \nu_{1,{\rm down}}\right) \epsilon_{2,{\rm up}} 
+  2  \sum_{- \Lambda}^{-1} \epsilon_{\rm up} + E_{\rm d.c.} .
\label{20}
\end{equation}
Up to an overall factor of 2 this is exactly the expression we would have to minimize when solving the one-flavor massive NJL model with coupling constant $g^2=2G^2$ (appearing in $E_{\rm d.c.}$)
and the valence
occupation fractions 
\begin{equation}
\nu_1 = \frac{1+ \nu_{1,{\rm up}}- \nu_{2,{\rm down}}}{2}, \quad \nu_2 = \frac{1+ \nu_{2,{\rm up}}- \nu_{1,{\rm down}}}{2}.
\label{21}
\end{equation}
Note that these effective occupation fractions correspond in general to excited NJL baryons with unoccupied states below occupied ones,
even if we are only interested in ground state baryons of the isoNJL model. 
Moreover, one can show that this procedure yields a self-consistent HF solution of the isoNJL model in terms of a solution of the NJL model.
This is important here since we have set ${\cal C}=0$ from the onset so that minimization of the energy does not guarantee self-consistency any more.

To demonstrate that we can really promote the NJL HF baryon to a isoNJL HF baryon, we write down  
the self-consistency condition of the one-flavor NJL system, denoting continuum eigenspinor components by $\psi_{1,2}$ and bound state
spinor components by $\psi_{1,2}^{(i)}$,
\begin{equation}
{\cal D}-m_0 = - 2 N_c g^2 \left(  \nu_1 \psi_1^{(1)*} \psi_2^{(1)} +   \nu_2 \psi_1^{(2)*} \psi_2^{(2)}  + \sum_{\epsilon = -\Lambda}^{-1} \psi_1^* \psi_2 \right).
\label{22}
\end{equation}
Inserting $\nu_1,\nu_2$ from (\ref{21}) and identifying $g^2=2G^2$, we can cast (\ref{22}) into the form
\begin{eqnarray}
{\cal D}- m_0 & = & - 2 N_c G^2 \left( \nu_{1,{\rm up}}  \psi_1^{(1)*} \psi_2^{(1)} +   \nu_{2,{\rm up}} \psi_1^{(2)*} \psi_2^{(2)}  + \sum_{\epsilon = -\Lambda}^{-1} \psi_1^* \psi_2 \right)
\nonumber \\  
& & -2 N_c G^2 \left( (1- \nu_{2,{\rm down}})  \psi_1^{(1)*} \psi_2^{(1)} +  (1- \nu_{1,{\rm down}}) \psi_1^{(2)*} \psi_2^{(2)}  + \sum_{\epsilon = -\Lambda}^{-1} \psi_1^* \psi_2 \right).
\label{23}
\end{eqnarray} 
We now introduce charge conjugate spinors (labeled by the energy eigenvalue $\epsilon$) as follows
\begin{equation}
\tilde{\psi}(\epsilon) = i \gamma_1 \psi(- \epsilon).
\label{24}
\end{equation}
Then the 2nd line of (\ref{23}) can be expressed as  
\begin{equation}
- 2 N_c G^2 \left( - (1-\nu_{2,{\rm down}}) \tilde{\psi}_2^{(2)*}\tilde{\psi}_1^{(2)} - (1- \nu_{1,{\rm down}}) \tilde{\psi}_2^{(1)*} \tilde{\psi}_1^{(1)} -  \sum_{\epsilon = 1}^{\Lambda} \tilde{\psi}_2^* \tilde{\psi_1} \right).
\label{25}
\end{equation}
The scalar and pseudoscalar condensates vanish if summed over all states, rather than over the occupied states,
\begin{equation}
0 = \tilde{\psi}_2^{(1)*}\tilde{\psi}_1^{(1)} + \tilde{\psi}_2^{(2)*}\tilde{\psi}_1^{(2)} + \sum_{\epsilon=-\Lambda}^{-1} \tilde{\psi}_2^*\tilde{\psi}_1 + \sum_{\epsilon=1}^{\Lambda} \tilde{\psi}_2^*\tilde{\psi}_1. 
\label{26}
\end{equation}
This follows from ${\rm Tr\,} \gamma^0$=0 and the completeness of the eigenstates of $H$. Eqs.~(\ref{25},\ref{26}) can be used to rewrite (\ref{23}) as
\begin{eqnarray}
{\cal D}- m_0 & = & - 2 N_c G^2 \left( \nu_{1,{\rm up}}  \psi_1^{(1)*} \psi_2^{(1)} +   \nu_{2,{\rm up}} \psi_1^{(2)*} \psi_2^{(2)}  + \sum_{\epsilon = -\Lambda}^{-1} \psi_1^* \psi_2 \right)
\nonumber \\  
& & -2 N_c G^2 \left(  \nu_{1,{\rm down}} \tilde{\psi}_2^{(1)*} \tilde{\psi}_1^{(1)} +  \nu_{2,{\rm down}} \tilde{\psi}_2^{(2)*} \tilde{\psi}_1^{(2)} +  \sum_{\epsilon = -\Lambda}^{-1} \tilde{\psi}_2^* \tilde{\psi_1} \right).
\label{27}
\end{eqnarray} 
But this is precisely the first line of the self-consistency relation (\ref{11}) if we use the isospin up and down spinors
\begin{equation}
 \left( \begin{array}{c} \Psi_{1,1} \\ \Psi_{1,2} \\ \Psi_{2,1} \\ \Psi_{2,2} \end{array} \right)_{\uparrow} = \left( \begin{array}{c} \psi_1 \\ 0 \\ \psi_2 \\ 0 \end{array} \right), 
\quad \left( \begin{array}{c} \Psi_{1,1} \\ \Psi_{1,2} \\ \Psi_{2,1} \\ \Psi_{2,2} \end{array} \right)_{\downarrow} =
 \left( \begin{array}{c} 0 \\ \tilde{\psi_1} \\ 0 \\ \tilde{\psi_2}  \end{array} \right) 
\label{28}
\end{equation}
as solutions of the Dirac-HF equation (\ref{12}).
The 2nd line of (\ref{11}) with ${\cal C}=0$ is trivially fulfilled, so that we have indeed self-consistency for baryons in the massive isoNJL model.

Let us now discuss the quantum numbers of the baryons. We define a reduced baryon number $n_B$ as fermion number divided by 2$N_c$ and a reduced isospin $n_3$  as the total isospin
$T_3$ divided by $N_c$. Both of these quantities are normalized such that they run from $ -1$ to $+1$ if the occupation fractions of the valence levels vary between 0 and 1.
They are related to the occupation fractions as
\begin{eqnarray}
n_B & = &  \frac{N_f}{2N_c} = \frac{\nu_{1,{\rm up}} + \nu_{2,{\rm up}} + \nu_{1,{\rm down}} + \nu_{2,{\rm down}} -2}{2},
\nonumber \\
n_3 & = & \frac{T_3}{N_c} = \frac{\nu_{1,{\rm up}} + \nu_{2,{\rm up}} - \nu_{1,{\rm down}} - \nu_{2,{\rm down}}}{2}.
\label{29}
\end{eqnarray}
We may actually restrict our attention to $n_B\ge 0, n_3\ge 0$. A change of sign of either $n_B$ or $n_3$ can be generated by isospin symmetry or particle-antiparticle symmetry (CPT invariance),
leaving the baryon mass unchanged. 

Let us concentrate on the lowest lying states for given baryon number and isospin, since excited states are unlikely to play any role
for the $T=0$ phase boundary. More precisely, we select the state with maximal isospin for given baryon number or, equivalently, the state of maximal baryon 
number for given isospin. This still leaves us with a continuous set of states in the large $N_c$ limit which we sample by a discrete set of 11 configurations, listed in 
Table~\ref{tab1}. For each set of (up and down) filling fractions in the isoNJL model, we indicate the reduced baryon number $n_B$, the reduced isospin $n_3$
and the effective occupation numbers $\nu_{1,2}$ in the equivalent one-flavor NJL baryon computation, see Eqs.~(\ref{21},\ref{29}). Since our binning is such that it would yield all possible states 
for $N_c=5$, we also show the quark configuration for this particular case in the last column, to guide the intuition. It goes without saying that the quark content
is different here from SU(5) QCD due to the lack of confinement.

The first column contains a label of the baryon states (1,...,11) for reference below. A few remarks are in order here.
\vskip 0.5cm
\begin{center}
\begin{table}
\begin{tabular}{|c|c|c|c|c|c|c|c|c|c |}
\hline
 & $\nu_{1,{\rm up}}$ & $\nu_{2,{\rm up}}$ & $\nu_{1,{\rm down}}$ & $\nu_{2,{\rm down}}$ & $n_B$ & $n_3$ & $\nu_1$ & $\nu_2$ & $N_c=5$ \\
\hline
1) &1 & 1 & 1 & 1 & 1 & 0 & .5 & .5 & $u^5 d^5$ \\
2) & 1 & 1 & 1 & .8 & .9 & .1 & .6 & .5 & $u^5 d^4$ \\
3) & 1 & 1 & 1 & .6 & .8 & .2 & .7 & .5 & $u^5 d^3 $ \\
4) & 1 & 1 & 1 & .4 & .7 & .3 & .8 & .5 & $u^5 d^2$ \\
5) & 1 & 1 & 1 & .2 & .6 & .4 & .9 & .5 & $u^5 d$ \\
6) & 1 & 1 & 1 & 0 & .5 & .5 & 1 & .5 & $u^5$ \\
7) & 1 & 1 & .8 & 0 & .4 & .6 & 1 & .6 & $u^5 \bar{d}$\\
8) & 1 & 1 & .6 & 0 & .3 & .7 & 1 & .7 & $u^5 \bar{d}^2$\\
9) & 1 & 1 & .4 & 0 & .2 & .8 & 1 & .8 & $u^5 \bar{d}^3$ \\
10) & 1 & 1 & .2 & 0 & .1 & .9 & 1 & .9 & $u^5\bar{d}^4$ \\
11) & 1 & 1 & 0 & 0 & 0 & 1 & 1 & 1 & $u^5 \bar{d}^5$ \\ 
\hline
\end{tabular}
\caption{Representative selection of isoNJL model baryons entering the construction of the $T=0$ phase boundary in the first quadrant of the ($\mu,\nu$)-plane. See main text for the explanation of the entries and a discussion.}
\label{tab1}
\end{table}
\end{center}
\begin{itemize}
\item  
In baryon 1), the isospin up and down levels are completely filled. This amounts to maximal baryon number and vanishing isospin. As shown in the table, it is equivalent to the one-flavor NJL baryon where each
level is half-filled. On the other hand, since the up- and down fillings are the same, we must have ${\cal D}={\cal D}^*$ or $P_3=0$. Hence it 
must also be equivalent to the GN baryon with maximal filling. Thus, the NJL baryon with 
$\nu_1=\nu_2=1/2$ is equivalent to a GN baryon, known analytically. 
\item
In baryon 11), the isospin up levels are completely filled, the down levels empty. This corresponds to a one-flavor NJL baryon and antibaryon and matches the mean fields ${\cal D},{\cal D}^*$. Hence this case
is identical to the baryon in the one-flavor NJL model with maximal filling. In the isoNJL model, it has vanishing baryon number and maximal isospin.
\item
Let us also discuss baryon 6) which will later turn out to play an important role for the phase diagram. The up levels are completely filled, the down level 1 is filled whereas 2 is empty.
From the point of view of the separate NJL baryons, it is a combination of a maximal baryon (up) and the vacuum (down). Seen from the isoNJL model this state
consists solely of the maximal number of up quarks. This leads to $n_B=n_3=1/2$ and an equivalent NJL baryon with half maximal fermion number. 
\end{itemize}
The most important message is the fact that all baryons of the massive isoNJL model can be reduced to baryons of the massive one-flavor NJL model. 
Since these have already been evaluated in Ref.~\cite{L18}, we can take over the techniques from \cite{L18} literally and refer to this paper for more 
details about the numerical HF calculation.
The only new aspect is the fact that we need excited NJL baryons as well, rather than the ground states
treated in Ref.~\cite{L18}, but this requires only a trivial change of the occupation fractions. 

\begin{figure}
\begin{center}
\epsfig{file=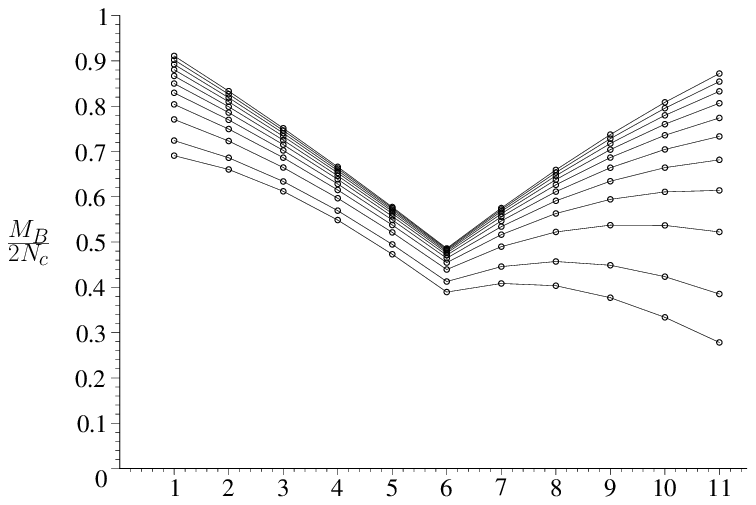,width=12cm,angle=0}
\caption{Baryon masses of the isoNJL model for $\gamma=0.05$ and 0.1,0.2,....,1.0, from bottom to top. The horizontal axis shows the labeling of states introduced in Table~\ref{tab1}. The points are calculated and connected
by straight line segments drawn to guide the eye. States with labels 1 (GN baryon), 6 (up baryon) and 11 (NJL baryon) play a prominent role in the phase diagram, but all points
are needed to construct the $T=0$ phase boundary.}
\label{fig2}
\end{center}
\end{figure}

We have evaluated the masses for all states of Table~\ref{tab1} for $\gamma=0.05$ and $\gamma=0.1,0.2,...,1.0$. The resulting
masses are shown in Fig.~\ref{fig2}. The numerically computed masses of baryons 1) (the leftmost points in Fig.~\ref{fig2}) can be compared to the analytically known GN masses
and agree within 4 digits. Likewise, the numerically and analytically computed order parameters are indistinguishable on a plot. This gives us confidence that the results shown in Fig.~\ref{fig2}
are sufficiently accurate. For the convenience of the reader, we 
quote here the formulas determining the baryon masses of the massive GN model \cite{L16,L17}. If the two valence states are occupied with fractions $\nu_1,\nu_2$, the mass is simply given by 
\begin{equation}
\frac{M_B}{N_c}  =  \frac{2}{\pi}\left[  \sin \theta + \gamma {\rm \, artanh}(\sin \theta) \right]
\label{30}
\end{equation}
where $\theta$ is the solution of the transcendental equation
\begin{equation}
\frac{1-\nu_1+\nu_2}{2} = \frac{1}{\pi}\left( \theta+\gamma \tan \theta \right).
\label{31}
\end{equation}
Please note that our present definition of $\gamma$, Eq.~(\ref{4}), differs from the one used in the original references by a factor of $\pi$.

%###########################################################################################################################
%####################################################################################################\#######################

%<<<<<<<<<<<<<<<<<<<<<<<<<<<<<<<<<<<<<<<<<<<<<<<<<<<<<<<<<<<<<<<<<<<<<<<<<<<<<<<<<<<<<<<<<<<<<<<<<<<<<<<<<<<<<<<<<<<<<<<<<<
%<<<<<<<<<<<<<<<<<<<<<<<<<<<<<<<<<<<<<<<<<<<<<<<<<<<<<<<<<<<<<<<<<<<<<<<<<<<<<<<<<<<<<<<<<<<<<<<<<<<<<<<<<<<<<<<<<<<<<<<<<<
\section{Phase boundaries at zero temperature}
\label{sect3}
%<<<<<<<<<<<<<<<<<<<<<<<<<<<<<<<<<<<<<<<<<<<<<<<<<<<<<<<<<<<<<<<<<<<<<<<<<<<<<<<<<<<<<<<<<<<<<<<<<<<<<<<<<<<<<<<<<<<<<<<<<<
%<<<<<<<<<<<<<<<<<<<<<<<<<<<<<<<<<<<<<<<<<<<<<<<<<<<<<<<<<<<<<<<<<<<<<<<<<<<<<<<<<<<<<<<<<<<<<<<<<<<<<<<<<<<<<<<<<<<<<<<<<<

In both the GN and the NJL models, the critical chemical potential $\mu_c$ is given by the mass of the baryon with maximal fermion number.
How does this generalize to the isoNJL model, where we expect a whole critical curve in the ($\mu,\nu$)-plane? All we know so far are the endpoints of this
curve on the $\mu$-axis (the mass of the GN baryon) and the $\nu$-axis (the mass of the NJL baryon). 

Let us first understand the relationship for the case of a single chemical potential in simple terms. Take a configuration of one baryon in the whole space 
(approximated by a finite box of length $L$ for this purpose). The grand canonical potential density at $T=0$ is given by
\begin{equation}
{\cal V}_{\rm eff} = {\cal E} - \mu \rho = \frac{M_B-\mu}{L}
\label{32}
\end{equation}
since the mean baryon density is $1/L$. Expression (\ref{32}) crosses zero at the point where $\mu=M_B$. Therefore we 
can identify $M_B$ with the critical chemical potential $\mu_c$ where the crystal phase starts. This has indeed been confirmed by previous studies of the massive GN 
and NJL models \cite{L6,L7}.

In the isoNJL model, we have the choice of bound states with a whole range of baryon and isospin numbers, see Table~\ref{tab1}.
As we can read off Table~\ref{tab1}, the relevant ones for us with positive baryon number and isospin have 
\begin{eqnarray}
\nu_{1,{\rm up}} & = & \nu_{2,{\rm up}} = 1,
\nonumber \\
\nu_{1,{\rm down}} & = & 1, \ \nu_{2,{\rm down}} =0...1 \quad {\rm or} \quad \nu_{1,{\rm down}}  =  0...1, \ \nu_{2,{\rm down}} =0.
\label{33}
\end{eqnarray} 
They carry the following reduced baryon number and isospin
\begin{equation}
n_B=  \frac{\nu_{1,{\rm down}}+ \nu_{2,{\rm down}}}{2}, \quad n_3 = 1 - n_B.
\label{34}
\end{equation}
Thus for example, the GN baryon (state 1 in Table~\ref{tab1}) has $\nu_{1,{\rm down}}=\nu_{2,{\rm down}}=1,n_B=1,n_3=0$. The NJL baryon 
(state 11) has
$\nu_{1,{\rm down}}=\nu_{2,{\rm down}}=0,n_B=0,n_3=1$. The up baryon (state 6) has $\nu_{1,{\rm down}}=1, \nu_{2,{\rm down}}=0,n_B=n_3=1/2$.
The mass of any baryon depends on baryon number $n_B$ and isospin $n_3$. 
Consider a state of one such baryon in the whole space with the effective potential
\begin{equation}
{\cal V}_{\rm eff} = \frac{M(n_B,n_3) - n_B \mu - n_3 \nu}{L} .
\label{35}
\end{equation}
The critical potentials $\mu_c,\nu_c$ where ${\cal V}_{\rm eff}$ changes sign must satisfy
\begin{equation}
M(n_B,n_3) - n_B \mu_c - n_3 \nu_c = 0 .
\label{36}
\end{equation}
This defines a straight line in the ($\mu,\nu$)-plane. If we move away from the origin of this plane and cross the critical line, it becomes 
advantageous for the low density system to start crystallization, using the structure dictated by the single baryon.
Since there are many different baryons, we can construct many different straight lines, using as input the baryon masses.
Only by considering all possible critical straight lines will we be able to identify the true critical curve in the ($\mu,\nu$)-plane.

\begin{figure}
\begin{center}
\epsfig{file=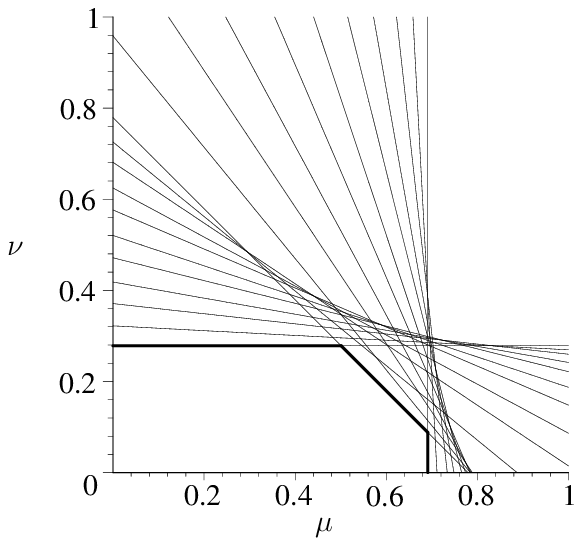,width=6cm,angle=0}\epsfig{file=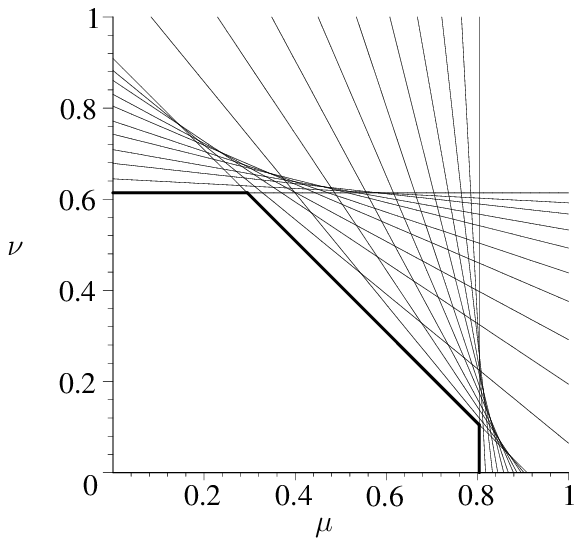,width=6cm,angle=0}\epsfig{file=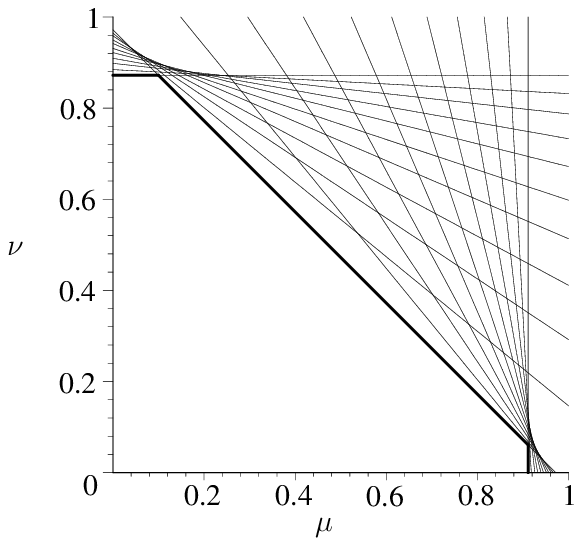,width=6cm,angle=0}
\caption{Families of critical straight lines according to Eq.~(\ref{36}) for $\gamma=0.05,0.3,1.0$ (from left to right). The true phase boundary emerges
by keeping only the resulting polygon closest to the origin. Only 3 different straight lines contribute, corresponding to the masses of
GN, NJL, and up baryons, see main text. This result holds independently of $\gamma$.} 
\label{fig3}
\end{center}
\end{figure}

Fig.~\ref{fig3} shows three examples of such a construction for three different values of the bare mass parameter $\gamma$. We have taken
into account all 11 states from Table~\ref{tab1} and Fig.~\ref{fig2}, plus 10 interpolated mass values half way between two adjacent points
in Fig.~\ref{fig2}. These interpolated straight lines have been included so as to check the stability of the final phase boundary 
against refining the discretization used in Table~\ref{tab1}. The outcome, valid also for the other values of $\gamma$ not shown here,  
is surprisingly simple.  

%###########################################################################################################################
%####################################################################################################\#######################

Common to all cases is the fact that only 3 baryons from Fig.~\ref{fig2} contribute to the final phase boundary: The GN baryon (1, leftmost), the NJL baryon (11, rightmost)
and the up baryon (6, midpoint). Since this depends on the details of the spectra shown in Fig.~\ref{fig2}, it is nothing that could have been anticipated on general grounds.
It implies that one needs only 3 baryon masses to construct the complete phase boundary at $T=0$, $M_{\rm GN}, M_{\rm NJL}$ and $M_{\rm up}$. The filling fractions
of the equivalent NJL baryon are $(\nu_1,\nu_2)= (1/2,1/2),(1,1),(1,1/2)$, respectively.
Since these masses play a key role in the phase diagram, we list them in Table~\ref{tab2} for the available $\gamma$-parameters.
Constructing the $T=0$ phase boundary in the ($\mu,\nu$)-plane with the help of Table~{\ref{tab2} is now straightforward. 
Draw the vertical line $\mu=M_{\rm GN}$, the horizontal line $\nu=M_{\rm NJL}$ and a straight line with slope $-1$ connecting the points $\mu=2 M_{\rm up}$ on the $\mu$-axis 
and $\nu= 2 M_{\rm up}$ on the $\nu$-axis. The slope $-1$ of the last straight line expresses the fact that the instability is driven by pure up quark bound states, presumably 
because the mass of these bound states lies in a deep, local minimum, see Fig.~\ref{fig2}. By reflections on both axes, the graph of the phase boundary may be extended
to the other quadrants of the chemical potential plane, yielding an octogon. This is shown in Fig.~\ref{fig4} for all $\gamma$ values used here. In the chiral limit,
the extension of the octogon in $\nu$-direction vanishes since the NJL baryon becomes massless. Consequently, the octogon degenerates to a segment of a line with endpoints
$\pm M_{\rm GN}(\gamma=0) = \pm 2/\pi$. It is also included in Fig~\ref{fig4}. In the opposite ``heavy quark" limit ($\gamma \to \infty$), all binding effects become negligible
and the relevant masses are $M_{\rm GN}=M_{\rm NJL}=1,M_{\rm up}=1/2$. The corresponding contour becomes a square shape connecting the points $\mu=\pm 1, \nu=\pm 1$ on
the coordinate axes, also included in Fig.~\ref{fig4} as the outermost curve.

The vertices of the octogon are actually triple points, since the homogeneous solution and two different crystal solutions meet there. Let us note the coordinates of these vertices: 
NJL and up intersect at $\mu=2M_{\rm up}-M_{\rm NJL}, \nu=M_{\rm NJL}$, GN and up intersect at  $\mu=M_{\rm GN}, \nu=2 M_{\rm up}-M_{\rm GN}$. As one crosses the critical curve and enters the
crystal phase, one expects that the shape of the baryons will survive at low densities, as for the one-flavor models. We would therefore expect further phase boundaries inside the crystal
phase starting from the triple points, but these cannot possibly be constructed using only baryon input.

%###########################################################################################################################
\begin{figure}
\begin{center}
\epsfig{file=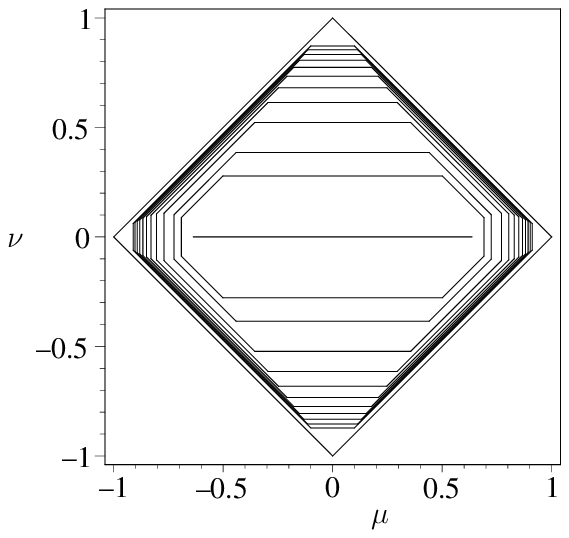,width=12cm,angle=0}
\caption{Phase boundaries of the massive isoNJL model at $T=0$, extended to the whole $(\mu,\nu$)-plane by using isospin and CPT symmetries. The octogons correspond to
$\gamma=0.05$ and 0.1,0.2,...,1.0, from inside out. Also shown is the chiral limit ($\gamma=0$) as a straight line segment in the center of the plot, and the heavy quark asymptote
($\gamma \to \infty$) as the outermost contour, a square.}
\label{fig4}
\end{center}
\end{figure}
%####################################################################################################\#######################

\vskip 0.5cm
\begin{center}
\begin{table}
\begin{tabular}{|c|c|c|c|}
\hline
$\gamma$ & 1) GN & 6) up  & 11) NJL \\
\hline
.05 & .6909 & .3896 & .2783 \\
.1 & .7240 & .4129 & .3853 \\
.2 & .7708 & .4393 & .5225 \\
.3 & .8041 & .4546 & .6142 \\
.4 & .8297 & .4644 & .6816 \\
.5 & .8501 & .4710 & .7334 \\
.6 & .8668 & .4758 & .7741 \\
.7 & .8807 & .4794 & .8066 \\
.8 & .8924 & .4821 & .8329 \\
.9 & .9025 & .4843 & .8544 \\
1. & .9112 & .4861 & .8721 \\
\hline
\end{tabular}
\caption{Results for isoNJL baryon masses divided by $2N_c$. These are all the values needed to construct the $T=0$ phase boundaries
shown in Fig.~\ref{fig4}.} 
\label{tab2}
\end{table}
\end{center}

%###########################################################################################################################
\begin{figure}
\begin{center}
\epsfig{file=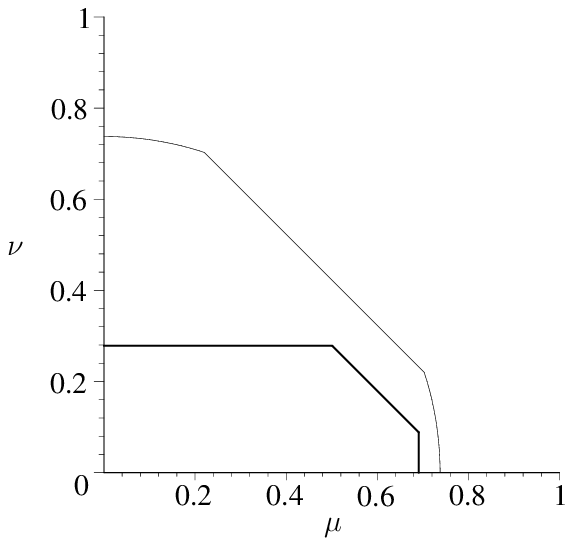,width=6cm,angle=0}\epsfig{file=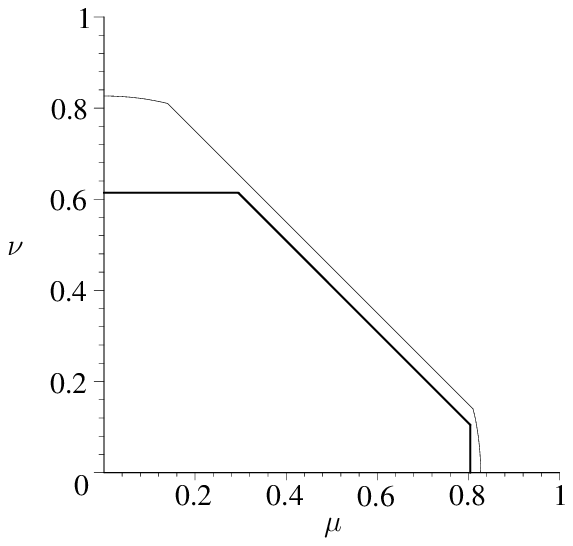,width=6cm,angle=0}\epsfig{file=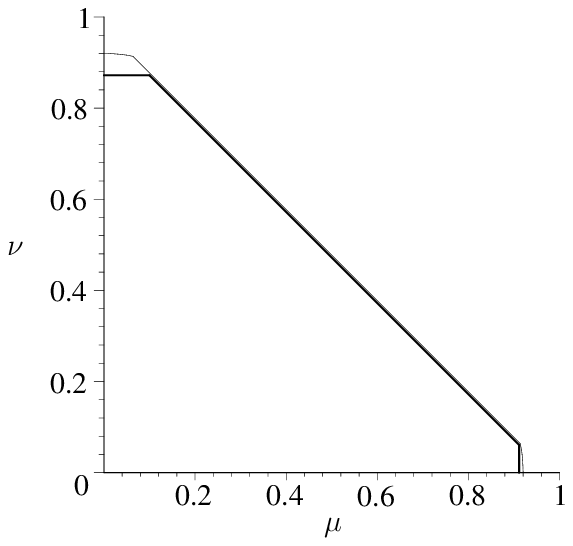,width=6cm,angle=0}
\caption{Comparison of $T=0$ first order phase boundaries from the full calculation (polygons, fat lines) and the homogeneous calculation (rounded off polygons, thin lines),
for $\gamma=0.05,0.3,1.0$ from left to right. The phase boundary according to the homogeneous calculation lies always inside the crystal phase and is therefore
invalidated.}
\label{fig5}
\end{center}
\end{figure}
%####################################################################################################\#######################

In the one-flavor NJL and GN models, the original calculation restricted to homogeneous phases also showed a first order transition at $T=0$ and a critical
chemical potential \cite{L19}. Here the fermion mass changes discontinuously. Since this critical potential was larger than the baryon mass, this alleged phase transition 
turned out to be irrelevant for the true phase diagram. 
Likewise, in order to fully validate our calculation of the critical curve in the isoNJL model, we have to check that the homogeneous calculation does not predict any first
order transition inside the contours shown in Fig.~\ref{fig4}. It is straightforward to determine the critical line in the ($\mu,\nu$)-plane where the first order
transition happens, assuming only homogeneity. A few representative results are shown in Fig.~\ref{fig5}. For all values of $\gamma$ considered in the present work,
the contour from the inhomogeneous calculation (using baryon masses as input) lies inside the contour from the homogeneous calculation. With increasing $\gamma$, the 
corresponding contours approach each other, but never cross. Clearly, the physical nature of the transition remains very different. In the limit $\gamma \to \infty$, the 
homogeneous phase boundary becomes the straight line joining the points $(\mu,\nu)=(0,1)$ and $(1,0)$ in a plot like Fig.~\ref{fig5}. As expected, this agrees perfectly
with the limiting curve in the whole ($\mu,\nu$)-plane discussed above and shown in Fig.~\ref{fig4}, underlining the consistency of the present isoNJL phase diagram. 

%<<<<<<<<<<<<<<<<<<<<<<<<<<<<<<<<<<<<<<<<<<<<<<<<<<<<<<<<<<<<<<<<<<<<<<<<<<<<<<<<<<<<<<<<<<<<<<<<<<<<<<<<<<<<<<<<<<<<<<<<<<
%<<<<<<<<<<<<<<<<<<<<<<<<<<<<<<<<<<<<<<<<<<<<<<<<<<<<<<<<<<<<<<<<<<<<<<<<<<<<<<<<<<<<<<<<<<<<<<<<<<<<<<<<<<<<<<<<<<<<<<<<<<
\section{Phase boundaries at finite temperature}
\label{sect4}
%<<<<<<<<<<<<<<<<<<<<<<<<<<<<<<<<<<<<<<<<<<<<<<<<<<<<<<<<<<<<<<<<<<<<<<<<<<<<<<<<<<<<<<<<<<<<<<<<<<<<<<<<<<<<<<<<<<<<<<<<<<
%<<<<<<<<<<<<<<<<<<<<<<<<<<<<<<<<<<<<<<<<<<<<<<<<<<<<<<<<<<<<<<<<<<<<<<<<<<<<<<<<<<<<<<<<<<<<<<<<<<<<<<<<<<<<<<<<<<<<<<<<<<

The perturbative phase boundary between homogeneous and inhomogeneous phases in ($\mu,\nu,T$)-space was the subject of Ref.~\cite{L15}. The last section
of the present work has equipped us with the nonperturbative phase boundary at $T=0$. What is still missing is the nonperturbative,
first order phase boundary bridging the gap between these two building blocks. We are not aware of any shortcut for such a calculation, therefore we have to resort to a full, numerical
HF calculation. This is obviously the most tedious part of the whole project. As in Ref.~\cite{L15}, we work with the grand canonical potential and introduce
chemical potentials conjugate to baryon ($\mu$) and isospin ($\nu$) densities. We once again assume that the charged pion condensate vanishes (${\cal C}=0$)
so that the Dirac-HF equations for the two isospin channels decouple,
\begin{eqnarray}
\left( \begin{array}{cc} i \partial_x -\mu-\nu & {\cal D}^* \\ {\cal D}  & - i \partial_x -\mu-\nu    \end{array} \right) 
\left( \begin{array}{c} \Psi_{1,1} \\ \Psi_{2,1} \end{array} \right) & = &  \omega \left( \begin{array}{c} \Psi_{1,1} \\ \Psi_{2,1} \end{array} \right) ,
\nonumber \\
\left( \begin{array}{cc} i \partial_x -\mu+\nu   & {\cal D} \\ {\cal D}^*  & - i \partial_x -\mu+\nu   \end{array} \right) 
\left( \begin{array}{c} \Psi_{1,2} \\ \Psi_{2,2} \end{array} \right) & = &  \omega \left( \begin{array}{c} \Psi_{1,2} \\ \Psi_{2,2} \end{array} \right) .
\label{37}
\end{eqnarray}  
Unlike at $T=0$, here we do not claim to find the fully self-consistent HF solution. Instead we perform a variational calculation, computing the grand canonical
potential and minimizing it with respect to the mean field ${\cal D}$. Since we tacitly assume that ${\cal C}=0$, self-consistency then does not follow automatically. 
At $T=0$ it was possible to prove that the minimum is nevertheless self-consistent, but here this is no longer the case. Both the chemical potentials and 
finite temperature invalidate the proof given above for baryons. We have to stick to the assumption ${\cal C}=0$ because of its enormous technical advantage. Indeed,
we see from Eq.~(\ref{37}) that the calculation of the grand canonical potential separates into isospin up and isospin down sectors, each one
being identical to the one-flavor NJL model. 
This allows us to use the whole apparatus developed for the massive NJL model from Ref.~\cite{L7}. The only new aspect is the fact that we have to run the part
of the calculation where the grand canonical potential is evaluated twice, using different mean fields (${\cal D},{\cal D}^*$) and different chemical potentials ($\mu \pm \nu$),
and then minimize the sum of the two results. Since the technicalities of such a thermal HF calculation have been described in all detail in Ref.~\cite{L7} and our approach here is literally the same,
we skip the details. The first order phase boundary is constructed as follows: For a given point in ($\mu,\nu,T$)-space, we evaluate the grand canonical
potential and minimize it with respect to 14 parameters (6 Fourier coefficients for each of the real mean fields $S$ and $P_3$, a constant mode $m$ and 
the period entering the Fourier series). For the minimization, we use a standard conjugate gradient algorithm. We choose trajectories in a plane $T={\rm const.}$, hopefully
intersecting the phase boundary, along which we compute the grand canonical potential and compare it with the homogeneous calculation. If the phase transition is of first order,
we can follow both solutions across the boundary and get a picture similar to the one shown in Fig.~8 of Ref.~\cite{L7}. If the transition is of second order, the 
inhomogeneous solution disappears at the phase boundary and the picture looks like Fig.~9 of Ref.~\cite{L7}. Since the calculation has to be repeated many times and we use
Maple on a desktop computer, this whole procedure is very time consuming. Therefore we have determined the finite temperature, first order phase boundary for 3 values of the mass parameter $\gamma$
only ($\gamma=0.1,0.3,0.5$).

The results are shown in Fig.~\ref{fig6}. These plots include the perturbative phase sheets from Ref.~\cite{L15} and the $T=0$ phase boundaries
from the previous section. The crosses are the numerical HF results. In almost all cases, we see a clear first order transition. A few points near the perturbative sheet 
are suggestive of a second order transition, but we cannot rule out a weak first order transition. Near the upper end of the first order sheet, it becomes more and more 
difficult to disentangle the homogeneous from the inhomogeneous solutions, since the values of the thermodynamic potentials get very close. 
At our present level of sophistication, we are not really able to locate the tricritical curve expected to separate first and second order transitions. 
By and large, the first order sheet follows closely the $T=0$ contour, steeply rising with increasing temperature. It is located exactly where it was to be expected.
This completes the determination of the phase boundary delimiting the inhomogeneous phase in ($\mu,\nu,T$)-space as a function of the bare mass parameter $\gamma$.
    
%###########################################################################################################################
\begin{figure}
\begin{center}
\epsfig{file=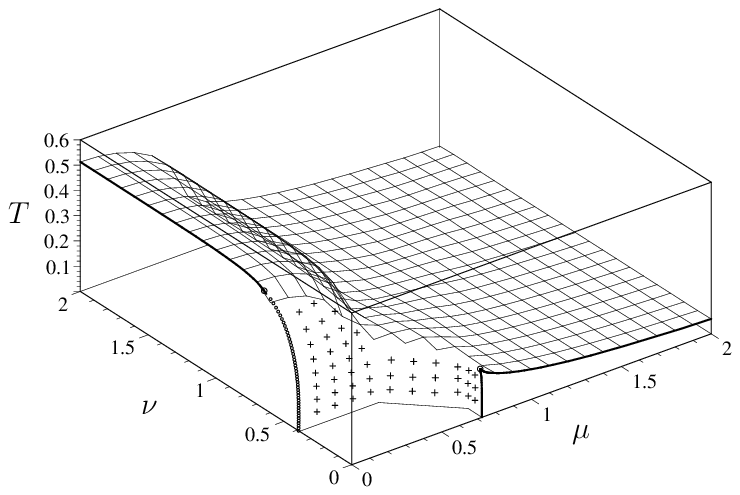,height=6cm,width=6cm,angle=0}\epsfig{file=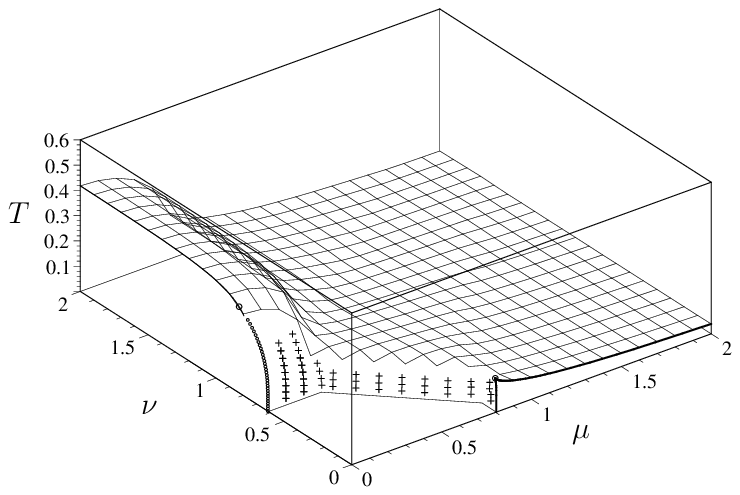,height=6cm,width=6cm,angle=0}\epsfig{file=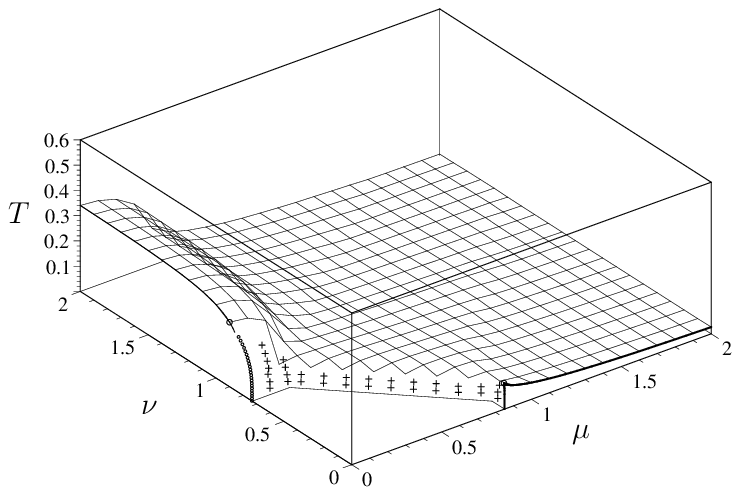,height=6cm,width=6cm,angle=0}
\caption{Full phase diagrams for the isoNJL model, showing the phase boundary sheet separating the inhomogeneous from the homogeneous phase, for 3 values of the bare mass ($\gamma=0.1,0.3,0.5$ from
left to right). The crosses lie on the first order sheet, joining the $T=0$ phase boundary with the perturbative sheet, and result from a numerical HF calculation
as described in the main text.} 
\label{fig6}
\end{center}
\end{figure}
%####################################################################################################\#######################

%<<<<<<<<<<<<<<<<<<<<<<<<<<<<<<<<<<<<<<<<<<<<<<<<<<<<<<<<<<<<<<<<<<<<<<<<<<<<<<<<<<<<<<<<<<<<<<<<<<<<<<<<<<<<<<<<<<<<<<<<<<
%<<<<<<<<<<<<<<<<<<<<<<<<<<<<<<<<<<<<<<<<<<<<<<<<<<<<<<<<<<<<<<<<<<<<<<<<<<<<<<<<<<<<<<<<<<<<<<<<<<<<<<<<<<<<<<<<<<<<<<<<<<
\section{Summary and conclusions}
\label{sect5}
%<<<<<<<<<<<<<<<<<<<<<<<<<<<<<<<<<<<<<<<<<<<<<<<<<<<<<<<<<<<<<<<<<<<<<<<<<<<<<<<<<<<<<<<<<<<<<<<<<<<<<<<<<<<<<<<<<<<<<<<<<<
%<<<<<<<<<<<<<<<<<<<<<<<<<<<<<<<<<<<<<<<<<<<<<<<<<<<<<<<<<<<<<<<<<<<<<<<<<<<<<<<<<<<<<<<<<<<<<<<<<<<<<<<<<<<<<<<<<<<<<<<<<<

We are interested in the phase diagram of the isoNJL model in two dimensions, in the large $N_c$ limit. While a candidate for the phase diagram in the chiral limit
is already known analytically, the massive model with a bare fermion mass and explicit chiral symmetry breaking has not yet been fully solved. 
In a previous work, the perturbative (second order) phase boundary separating a high temperature homogeneous phase from a low temperature inhomogeneous phase 
had already been constructed. In the present work, we have studied the more challenging nonperturbative problem of finding the missing first order phase 
boundaries via numerical calculations.

At zero temperature, the problem simplifies. It turns out that it is sufficient to know the spectrum of baryons with different isospin to construct the phase boundaries.  
To find the baryon masses, it is necessary to do numerical HF calculations, but these are somewhat easier than the
calcuations of inhomogeneous phases at finite temperature. The final results are strikingly simple, showing that only three different baryons seem to trigger
crystallization, namely the ones with reduced baryon number and isospin $(n_B,n_3)=(1,1),(1,1/2),(1/2,1/2)$. As we have demonstrated, these three baryons
in turn can be obtained from the one-flavor GN model (the baryon with maximal occupation) known analytically, or the one-flavor NJL model (the baryons with half occupation
and full occupation). Had we known this beforehand, it would not have been necessary to do any new baryon calculation at all since these three masses were already known.
However we first had to rule out contributions to the phase boundary from all the other states.

At finite temperature, there is no shortcut known so that we had to do a thermal HF calculation with inhomogeneous condensates. Owing to our assumption that the charged pion
condensate vanishes, the effort is roughly equivalent to doing a one-flavor NJL HF calculation twice. This can still be handled with Maple. The resulting first order sheet
shows no surprise, rising essentially vertically from the $T=0$ phase boundary if temperature is increased. The expected tricritical curve at the boundary between
first and second order sheets could not be determined reliably with our present numerical accuracy.

Two questions are left open by this investigation: First, what happens if we relax the assumption ${\cal C}=0$? And secondly, are there more phases inside the inhomogeneous region?

The assumption ${\cal C}=0$ is technically very crucial, both for Ref.~\cite{L15} and here, since it leads to a block diagonal form of the HF Hamiltonian and allows one
to use the full machinery developed for the one-flavor NJL model long ago. If one wanted to keep ${\cal D}$ and ${\cal C}$ at the same time, one would have to start all over
again. At $T=0$, since we could prove self-consistency of our HF solution explicitly, we know that we have found exact baryon states of the isoNJL model.
Only if a calculation with ${\cal D}\neq 0, {\cal C}\neq 0$ would yield another set of bound states with various isospins and smaller masses,
the phase diagram could change. This does not seem very likely. At finite temperature, we have not been able to prove self-consistency, but we also cannot rule it out.
In any case, the overall picture of the phase diagram seems to be fairly consistent and we would be surprised if many of its aspects would not become exact in the large $N_c$ limit.

The 2nd question was about the interior crystal region of the phase diagram. As we have discussed, it is quite plausible that the vertices of the octogon shaped phase boundaries
at $T=0$ are triple points from where additional first order lines could emanate, reaching into the crystal region. Since we know from the chiral limit that the shape of the 
inhomogeneous order parameter can be very complicated, one would need a large number of partial waves and an extremely efficient
minimization algorithm to discriminate between different such crystal structures. This is unfortunately beyond our present reach.

%#############################################################################################################
\section*{Acknowledgement}
%#############################################################################################################

The author thanks Michael Buballa, Dirk Rischke and Marc Wagner for lively and stimulating discussions.

\end{document}